\documentclass[twocolumn, letter]{emulateapj}
\usepackage{graphics,graphicx}
\usepackage{color}
\usepackage{natbib}
\usepackage{hyperref}
\DeclareGraphicsExtensions{.png,.pdf,.jpg}

\newcommand{\fermi}{{\it Fermi}}

\newcommand{\phots}{ph cm$^{-2}$ s$^{-1}$}
\newcommand{\dg}{$^\circ$}
\newcommand{\g}{$\gamma$}

\newcommand{\er}{$\pm$}
\newcommand{\pkssev}{PKS\,1718$-$649}

\shorttitle{Gamma-ray detection of the CSO PKS\,1718$-$649 with \fermi-LAT}
\shortauthors{Migliori et al.}

\begin{document}
\title{First detection in gamma-rays of a young radio galaxy: Fermi-LAT observations of the Compact Symmetric Object PKS\,1718$-$649}

\author{G. Migliori\altaffilmark{1}, A. Siemiginowska\altaffilmark{2}, M. Sobolewska\altaffilmark{3,2}, A. Loh\altaffilmark{1}, S. Corbel\altaffilmark{1,4}, L. Ostorero\altaffilmark{5}, {\L}. Stawarz\altaffilmark{6}}
\altaffiltext{1}{Laboratoire AIM (CEA/IRFU - CNRS/INSU - Universit\'e Paris Diderot), CEA DSM/SAp, F-91191 Gif-sur-Yvette, France.}
\altaffiltext{2}{Harvard-Smithsonian Center for Astrophysics, 60 Garden St., Cambridge, MA 02138, USA.}
\altaffiltext{3}{Nicolaus Copernicus Astronomical Center, Bartycka 18, 00-716 Warsaw, Poland.}
\altaffiltext{4}{Station de Radioastronomie de Nan\c{c}ay, Observatoire de Paris, CNRS/INSU, Univ. Orl\'eans, F-18330 Nan\c{c}ay, France.}
\altaffiltext{5}{Dipartimento di Fisica, Universit\`a degli Studi di Torino and Istituto Nazionale di Fisica Nucleare (INFN), Via P. Giuria 1, 10125 Torino, Italy.}
\altaffiltext{6}{Astronomical Observatory, Jagiellonian University, ul. Orla 171, 30-244 Krak\'ow, Poland.}
\email{giulia.migliori@cea.fr}

\begin{abstract}
We report the $\gamma$-ray detection of a young radio galaxy, PKS\,1718$-$649, belonging
to the class of Compact Symmetric Objects (CSOs), with the Large Area Telescope (LAT) on board
the {\it Fermi} satellite. 
The third {\it Fermi} Gamma-ray LAT catalog
(3FGL) includes an unassociated $\gamma$-ray source, 3FGL\,J1728.0$-$6446, located close to PKS\,1718$-$649. Using the latest Pass 8 calibration,
we confirm that the best fit $1 \sigma$ position of the $\gamma$-ray source is compatible with the radio location of PKS\,1718$-$649.
Cross-matching of the $\gamma$-ray source position with the positions of  blazar sources from several catalogs yields negative
results.  Thus, we conclude that
PKS\,1718$-$649 is the most likely counterpart to the unassociated LAT source.
We obtain a detection test statistics TS$\sim 36$ ($>$5$\sigma$) with  a best fit photon spectral index  $\Gamma=$2.9\er0.3 and a 0.1-100 GeV photon flux density $F_{\rm 0.1-100GeV}=$(11.5\er0.3)$\times{\rm 10^{-9}}$ ph cm$^{-2}$ s$^{-1}$. 
We argue that the linear size ($\sim$2 pc), the kinematic age ($\sim$100 years), and the source distance ($z=0.014$) make PKS\,1718$-$649 an ideal candidate for \g-ray detection in the framework of the model proposing that the most compact and the youngest CSOs can efficiently produce GeV radiation via inverse-Compton scattering of  the ambient photon fields by the radio lobe non-thermal electrons.
Thus, our detection of the source in \g-rays establishes young radio galaxies as a distinct class of extragalactic high-energy emitters, and yields a unique insight on the physical conditions in compact radio lobes interacting with the interstellar medium of the host galaxy.

\end{abstract}


\section{Introduction}
The radio loud active galactic nuclei (RL AGN) constitute nearly 60\%
of all the \g-ray sources detected in the first four years of all-sky-survey of the  Large Area Telescope (LAT) on board the {\it Fermi} satellite, and listed in the Third {\it Fermi} Gamma-ray LAT catalog (3FGL; Acero et al. 2015).
The overwhelming majority (98\%) of 
the RL AGN  in the 3FGL have been classified as blazars, i.e. AGN with relativistic jets oriented
close to the line of sight of the observer (see the Third LAT AGN Catalog, 3LAC;
Ackerman et al. 2015). The remaining 2\% of the LAT AGN includes mostly
radio galaxies whose jets point away from the observer, the so-called misaligned AGN (MAGN, Abdo et al. 2010a), and a few AGN of other types (see Massaro, Thompson \& Ferrara 2016 for a review).
About one third of the 3FGL sources is still unidentified (1010 over 3033) and it is among them that we may expect to discover new classes of \g-ray emitters.

In blazars, Doppler boosted \g-ray emission is produced in a compact, relativistically moving jet region close to the AGN. The origin of the emission detected by \fermi-LAT in MAGN is less clear (Abdo et al. 2010a, Kataoka et al. 2011), and models of stratified jets or extended emitting regions have been considered.
So far, given the Point Spread Function (PSF) of \fermi-LAT, Centaurus A remains the only radio galaxy in the 3FGL associated with a $\gamma$-ray component that clearly extends beyond the central
nuclear region and contributes more than $\sim$50\% of
the total $\gamma$-ray flux (Abdo et al. 2010b). This
extended $\gamma$-ray emission spatially coincides with the giant radio lobes of Centaurus A and it is consistent with inverse-Comptonization (IC) of the relic cosmic microwave background (CMB) radiation.
In Centaurus A, the detection and imaging of the diffuse \g-ray emission, which is isotropic and not boosted by relativistic effects, is likely possible because of the   source physical extension ($\sim 600$\,kpc) and proximity (distance of $\sim 3.6$\,Mpc). 

Compact radio sources with radio structures fully contained within the
central regions of their host galaxies ($<1$\,kpc) are thought to be the progenitors of the large-scale radio galaxies (lobes' linear sizes of 10s--100s kpc). 
The initial stage of the radio source growth is represented by Compact Symmetric Objects (CSOs) with sub-kpc-scale structures, symmetric radio morphologies, total radio emission dominated by the mini-lobes, and kinematical ages smaller than a few thousand years
(review by Orienti 2016). CSOs belong to the spectral class of gigahertz-peaked spectrum (GPS) radio galaxies, characterized by a stable, convex radio spectrum peaking at GHz frequencies (O'Dea 1998 for a review). 
Some theoretical models predict that
CSOs should be relatively strong \g-ray
emitters as their compact radio lobes contain copious amount of highly
relativistic particles and are embedded in an environment rich in low-energy photons (see Stawarz at al. 2008, Ostorero et al. 2010). 
During the initial phase of expansion, the lobes could be the sites of production of bremsstrahlung \g-ray emission potentially detectable by \fermi-LAT (Kino et al. 2009).  
The radio lobes are expanding at sub-relativistic velocities, therefore the \g-ray emission is expected to be isotropic and steady on short timescales (up to months/years).
However, detections of the \g-ray emission
from young radio sources have been so far elusive even with {\it Fermi} (Migliori et al. 2014, D'Ammando et al. 2016). 

In this paper, we report for
the first time the detection with {\it Fermi}-LAT of \g-ray emission from the nearby ($z=0.014$) GPS radio galaxy \pkssev~(Tingay et al. 2015 and references therein). The double-lobed, compact ($\sim$2 pc) radio morphology of \pkssev~and its estimated age  ($\rm t_{age} \sim 100$ years, Giroletti \& Polatidis 2009) make this source one of the
youngest in the class of 
CSOs.    
Throughout the paper we assume a $\Lambda$CDM cosmology with $H_0=70$ km s$^{-1}$ Mpc$^{-1}$, $\Omega_M=$0.3 and $\Omega_{\Lambda}=$0.7.

\section{Observations}
\begin{figure}
\centering
\includegraphics[scale=0.28]{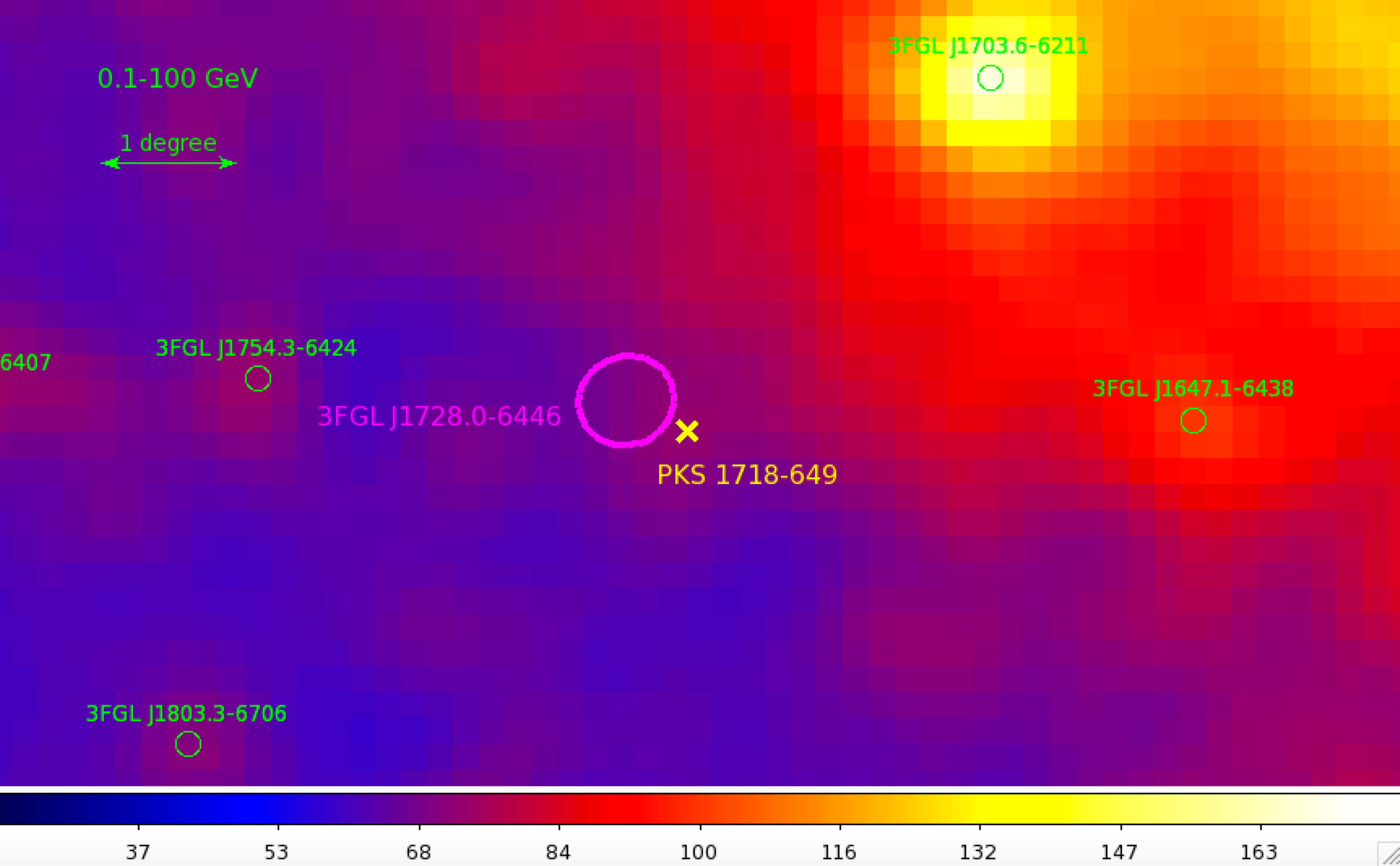}
\includegraphics[scale=0.34]{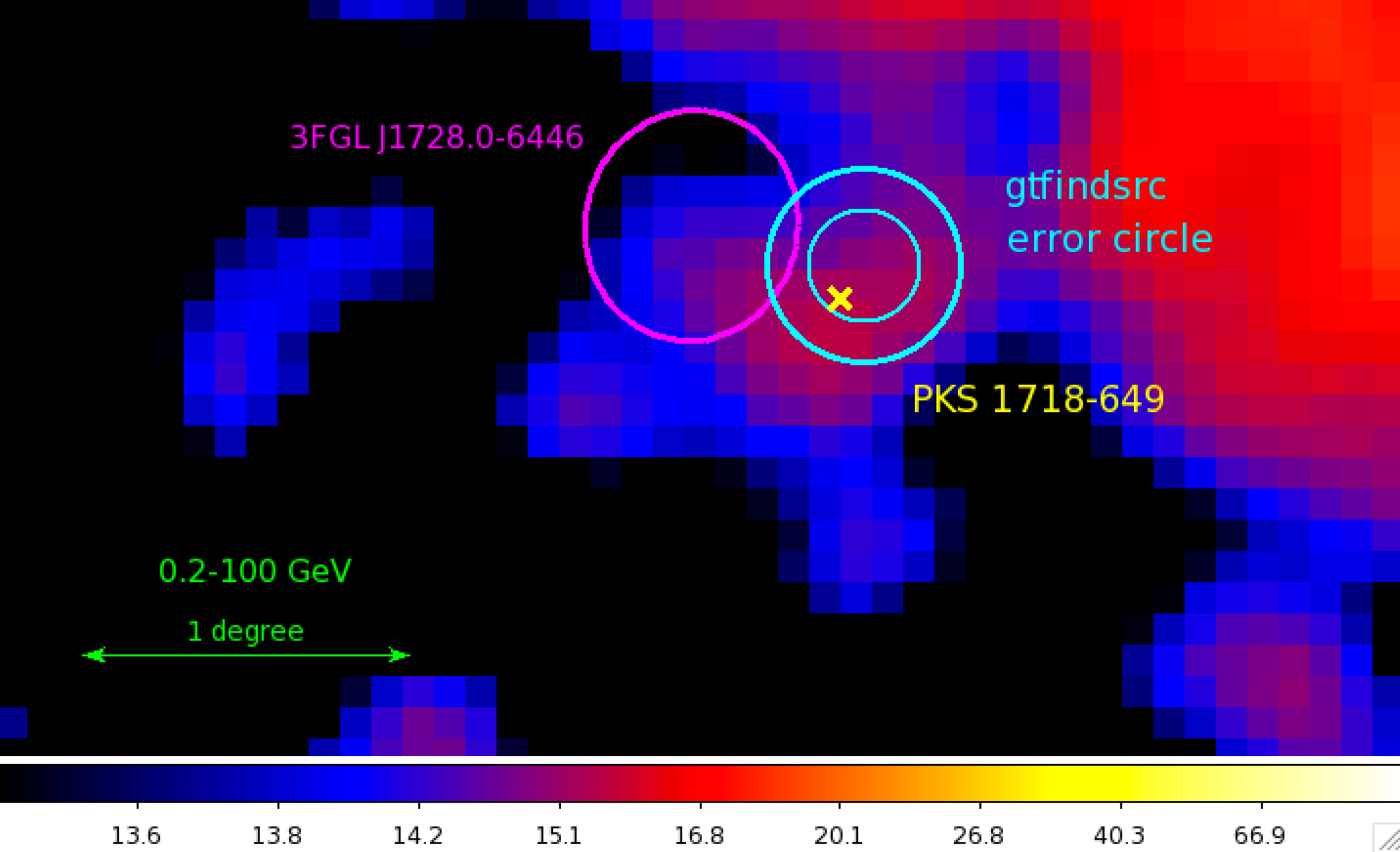}
\caption{Left: {\it Fermi}-LAT 0.1-100 GeV count map of the sky centered on the radio position of \pkssev~(yellow cross). The pixel size is 0.2 degrees/pixel. The magenta ellipse gives the uncertainty position (95\% confidence level) of 3FGL\,J1728.0$-$6446 reported in the 3FGL. Field \g-ray sources in the 3FGL are indicated with green circles. Right: Zoom on the \pkssev~region at $>$200 MeV energies. The pixel size is 0.1 degrees/pixel. The cyan circles correspond to the {\it gtfindsrc} best fit position (68\% and 95\% confidence levels).   }
\label{f1}
\end{figure}

\pkssev~belongs to the CSO sample selected for a \g-ray study with {\it Fermi}-LAT (Migliori 2016). 
This initial analysis of five year accumulation of the  {\it Fermi}-LAT data resulted in a marginal detection ($\approx$4$\sigma$) of a faint \g-ray source  at the radio position of \pkssev, F$_{\rm 0.1-100GeV}=$(1.5\er0.7)$\times 10^{-8}$ \phots.
The 3FGL contains an unassociated \g-ray source  (3FGL J1728.0-6446) detected at 4.4$\sigma$ level with a slightly different location, for which  \pkssev~lies just outside the 95\% uncertainty radius (Figure \ref{f1}, left panel, Table \ref{t2}). We thus performed a new analysis to assess the significance, flux, and location of the \g-ray source, exploiting the improved data quality and statistics of the new Pass 8 data release.

We analyzed seven years of the {\it Fermi}-LAT data (observations from
August 8, 2008 to August 8, 2015). We selected events of the Pass 8 (P8R2)
 Source class (Atwood et al. 2013) and used the adequate instrument response
functions for the analysis (P8R2\_V6) and the Science Tools
software package version v10r0p5.
The standard event selections were applied: we used FRONT and BACK events and applied a zenith angle cut at 90\dg~to eliminate Earth limb events.
We made use of both the binned and unbinned analyses. Given the long computational times due to the large dataset, we limited the unbinned analysis to consistency checks and, most importantly, to secure the source localization.

We first considered the 0.2$-$100 GeV band to minimize the systematic errors and study the background contamination (Ackermann et al. 2012). 
Next, we  increased the photon statistics by lowering the minimum energy threshold to 0.1 GeV. In the latter case, we excluded the data  with the lowest quality of the reconstructed direction\footnote{\url{http://fermi.gsfc.nasa.gov/ssc/data/analysis/documentation/Cicerone/Cicerone_Data/LAT_DP.html}.} (evtype=56) and applied a zenith angle cut at 80\dg, to minimize the contribution of the background \g-rays from the Earth's limb. 
We selected a circular region of interest
(RoI) of 15$^\circ$ radius, centered on the radio position of PKS 1718$-$649.  

The \fermi~source model that we adopted to calculate the
binned likelihood includes all the 3FGL point-like and diffuse sources
within the RoI. 
Additionally, we included the 3FGL sources falling between 15$^\circ$ and 25$^\circ$. In fact, due to the energy-dependent size of the {\it Fermi}-LAT PSF, these sources can contribute to the total counts observed within the RoI.
The 3FGL only accounts for sources detected within the first four years of observations. Therefore, we carefully applied this initial model to the seven-year dataset and, whenever necessary, we improved it by, e.g., adding sources that were detected in the following three years. 
In order to assess the fit quality, we inspected the residual maps obtained by comparing the {\it Fermi}-LAT count map with a count map created from the best-fit model. 
For example, we added a variable radio source, PKS~1824$-$582, located $\sim$10.3\dg~from the RoI center (flaring in April 2014, ATel \#6067, ATel \#6076).
The photon spectral indices, $\Gamma$, of the faintest sources within 7\dg~from the RoI center were fixed to the values reported in the 3FGL.
For sources located 7\dg~away from the RoI center, all the spectral parameters were initially fixed to the respective 3FGL values. We opportunely modified these values, whenever the inspection of the residuals revealed a significant variation of a source (e.g., 3FGL J1703.6-6211, ATel \#7330).  

The emission model also
accounts for the Galactic and extragalactic (and instrumental) diffuse
backgrounds. We used the `mapcube' file
\texttt{gll\_iem\_v06.fits} and the \texttt{iso\_P8R2\_V6\_v06.txt}
table to describe the emission from the Milky Way and the isotropic
component, respectively.

To test the hypothesis that the \g-ray source reported in the 3FGL is associated with \pkssev,  in the model we substituted 3FGL J1728.0-6446 with a source located at the radio position of \pkssev, assuming a power-law spectrum model  ($F=KE^{-\Gamma}$).
 We performed a fit of the data using the binned maximum likelihood ({\it gtlike}). The temporal behavior of the source was investigated with a lightcurve with one-year time resolution, and  in an incremental way, i.e. by progressively summing the years of observation.
 Quoted errors are at $1 \sigma$ statistical significance, if not otherwise specified.
 
\section{Results}
The results of the binned likelihood analysis performed on the seven-year dataset for the two low-energy  cutoffs are shown in Table \ref{t1}.
The binned analysis of the 0.2$-$100 GeV dataset yielded \g-ray emission detected at the \pkssev~position with a test significance\footnote{The test statistic is the logarithmic ratio of the
  likelihood of a source being at a given position in a grid to the
  likelihood of the model without the source,
  TS=2log(likelihood$_{src}$/likelihood$_{null}$) (Mattox et al. 1996).} TS$=$18.5 ($\sigma\sim$4.3), which improves over the TS=14 of our previous five-years-data analysis (Migliori 2016). 
We obtained an integrated source photon flux density above 0.2 GeV  $F_{\rm 0.2-100GeV}=$(2.1\er0.8)$\times 10^{-9}$ \phots~and a best fit photon spectral index  $\Gamma=$2.6\er0.3.  

The results improved when the analysis was extended to the 0.1-100 GeV band. We obtained a TS$\sim$36 detection ($\gtrsim 5\sigma$) at the location of \pkssev. The best fit photon spectral index, $\Gamma=$2.9\er0.3, is steeper than that measured at $\geq 0.2$ GeV, but the two values are consistent within the uncertainties. As a check, we restricted the analysis to the two best quality quartiles of the data in the PSF type partition (event types PSF2 and PSF3), i.e. the data with the best quality of the reconstructed direction of the photons. We used the Python package {\texttt SummedLikelihood} to combine the likelihoods of the separately analyzed (PSF2 and PSF3 type) datasets. Despite the significant cut in event number, we detected a source at the position of \pkssev~with a TS$\sim$24 and spectral parameters consistent within the uncertainties with the 0.1$-$100 GeV best fit values. 

The source is not detected at a significance level exceeding TS=10 on yearly timescales. The incremental time analysis shows that the source begins to be significantly detected over 4 years. The TS value progressively increases over 5, 6 and 7 years, whereas the measured flux does not vary significantly (Table \ref{t1}).

\subsection{Gamma-ray source Association}
In order to optimize the localization of the $\gamma$-ray source,
we ran the {\it gtfindsrc} tool, which calculates the best TS for different
positions given an initial guess (in this case \pkssev~coordinates) until the convergence tolerance for a positional fit is reached. 
The best fit position and the corresponding 68\% and 95\% error radii  are reported in Table \ref{t2}. \pkssev~lies within the 68\% confidence radius, at a distance of 0.13\dg~from the best fit \g-ray position (Figure \ref{f1}, right panel). 
We evaluated the significance of  a \g-ray source located at the position of 3FGL J1728.0-6446 using the 0.1-100 GeV and the 0.2-100 GeV datasets, and we obtained lower TS values ($<4\sigma$) than for \pkssev. Finally, we fit to the data a model with two distinct emitters located at the positions of \pkssev~and 3FGL J1728.0-6446: the \g-ray emission in excess over the background is attributed to the former, whereas the latter has TS$\sim$0.

We searched for other counterparts to the detected \g-ray source in catalogs and surveys of blazars and extragalactic radio sources, including the Roma BZCAT (Massaro et al. 2009),  a catalog of \g-ray candidates among WISE sources  (D'Abrusco et al. 2014), the Combined Radio All-Sky Targeted Eight GHz Survey (CRATES, Healey et al. 2009), the Parkes Catalog (Wright \& Otrupcek 1990), and the Parkes-MIT-NRAO survey (PMN, Gregory et al. 1994), and found no match within the {\it gtfindsrc} 95\% error circle. At low radio frequencies (843 MHz), a query to the Sydney University Molonglo Sky Survey (SUMSS, Mauch et al. 2003) returned 15 radio sources within the error circle.  
However, \pkssev~ is: 1) the brightest source at 843 MHz, with a flux density  $F_{\rm 843MHz}=($3.7\er0.1) Jy; 2) the only source reported in the high radio frequency catalogs, with a detection at 5.0 and 8.4 GHz, and a counterpart in the Two Micron All Sky Survey (2MASS, Skrutskie et al. 2006). For the remaining 14 SUMSS sources, the flux faintness (only three of them have 843 MHz fluxes in the 100--200 mJy range), non-detection at the higher frequencies, and lack of a 2MASS counterpart strongly disfavor the identification with an extragalactic, flat spectrum radio source. We concluded that there are no evident blazar candidates in the  95\% error circle of the \g-ray source.

\section{Discussion and Conclusions}
Our analysis of the seven-year {\it Fermi}-LAT data resulted in a $>$5$\sigma$ detection of the \g-ray source 3FGL J1728.0$-$6446, which had no association in the 3FGL catalog. Based on our revised best fit \g-ray position and the analysis of nearby sources, we concluded that this source is most likely the \g-ray counterpart to \pkssev, a young radio galaxy classified as  CSO. 
This represents the first significant ($>$5$\sigma$) \g-ray detection of a bona fide CSO, and it may provide us with important insights on the nature of the high-energy emission observed in young radio sources. 

\begin{figure}
\centering
\includegraphics[scale=0.455]{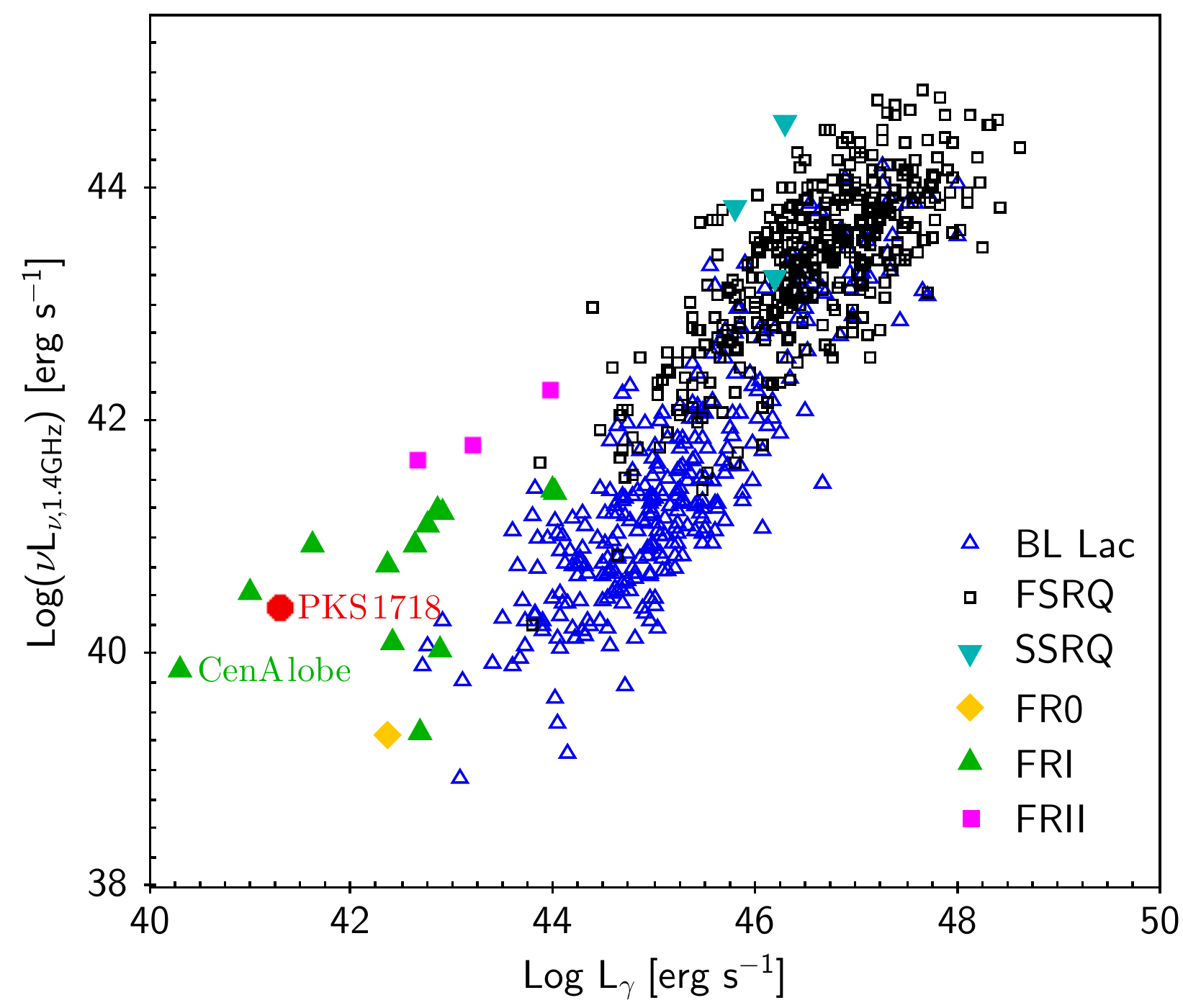}
\includegraphics[scale=0.45]{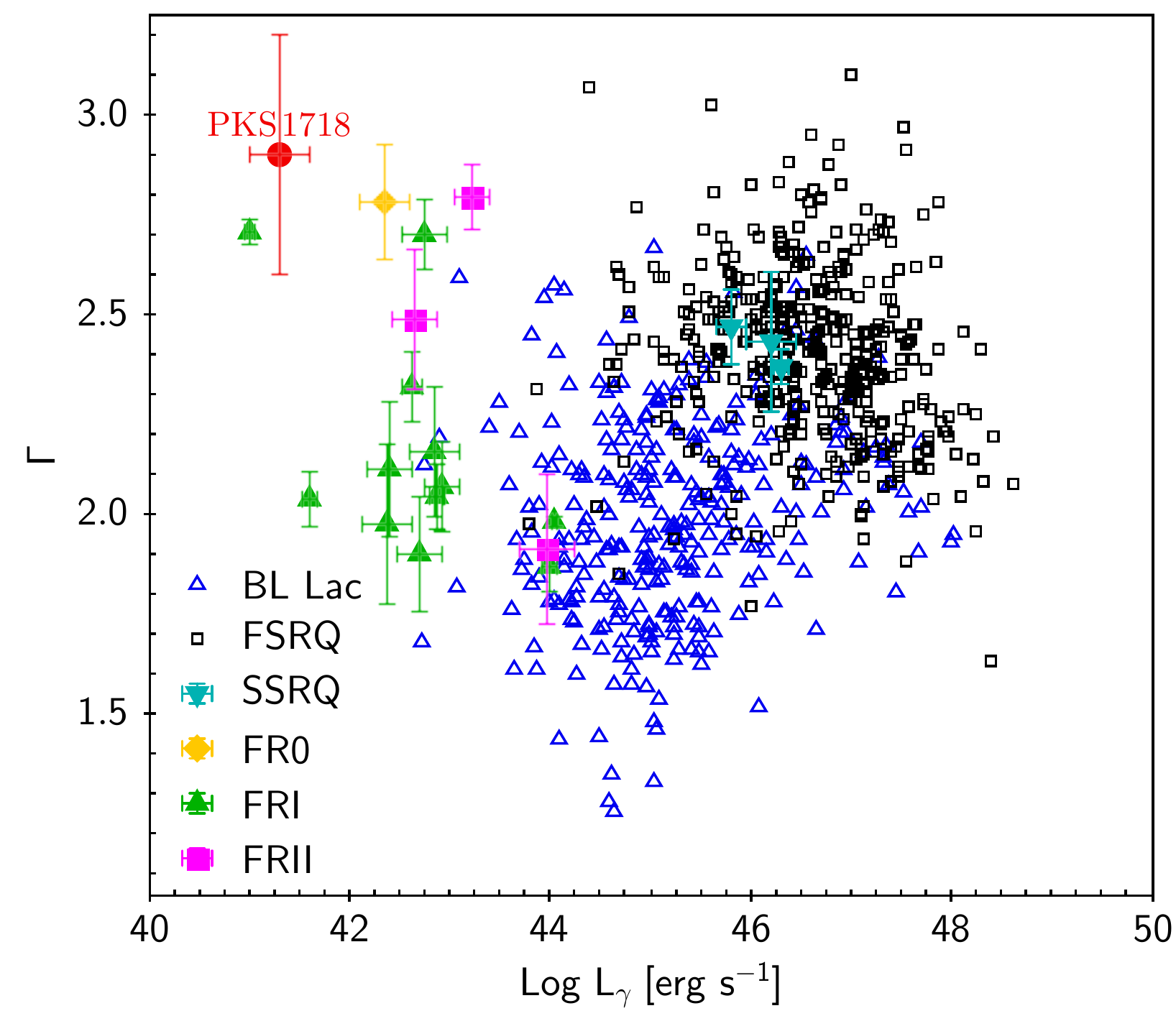}
\caption{ Left: Radio luminosity of the 3LAC FR I (solid, green triangles) and FR II (solid, magenta squares) radio galaxies,  BL Lacs (open, open triangles), FSRQs (open, black squares), and SSRQs (cyan, solid triangles) plotted as a function of the \g-ray luminosity between 1 GeV and 100 GeV. We included Tol1326-379 (solid, yellow diamond), classified as FR 0 (Grandi et al. 2015), and the \g-ray detected lobes of Centaurus A (the south lobe
with measured \g-ray and 1.4 GHz fluxes, see Abdo et al. 2010b and Hardcastle et al. 2009, respectively). The position of \pkssev~is indicated by the solid, red circle. Right: \g-ray spectral index versus 1-100 GeV luminosity. \pkssev~is located in the MAGN region of the diagram, with values of $\Gamma$ and $L_\gamma$ similar to those of Centaurus A. (Sample data: courtesy of P. Grandi.) }
\label{f2}
\end{figure}

We compared the \g-ray properties of \pkssev~with those of blazars and MAGN from the clean 3LAC sample. We considered only blazars with known redshift, and classified as either Flat Spectrum Radio Quasars (FSRQs) or BL Lac objects. The MAGN sample includes steep spectrum radio quasars (SSRQs), Fanaroff-Riley (FR) type I and type II radio galaxies (Fanaroff \& Riley 1974) detected by \fermi-LAT\footnote{For completeness we added 3C~111 (Grandi et al. 2012), and Cen~B (Katsuta et al. 2013), excluded from the clean 3LAC sample because of their low Galactic latitudes, and 3C~120 (Tanaka et al. 2014, Casadio et al. 2015).}, and Tol1326-379, the only FR~0 radio galaxy\footnote{FR~0 
have radio linear sizes LS$\lesssim$10 kpc, radio core powers similar to those of FR~Is, and a strong deficit of the corresponding extended radio emission (Baldi, Capetti \& Giovannini 2015).}  associated with a \g-ray source (Grandi et al. 2015). Figure \ref{f2} shows the position occupied by \pkssev~in the radio (1.4 GHz rest frame) vs. \g-ray (integrated above 1 GeV) luminosity plot (left panel). The 1.4 GHz luminosity of \pkssev~was calculated from the ATCA flux, $F_{\rm 1.4GHz}=$3.98 Jy (Maccagni et al. 2014). The source is located in the low radio and \g-ray luminosity region occupied by MAGN. 

The position of \pkssev~in the $\Gamma$ vs. L$_{\gamma}$ plot also supports a non-blazar origin of its  \g-ray emission: the source occupies the upper left corner, well separated from the blazar sources (Figure \ref{f2}, right panel). It is interesting to note that its \g-ray luminosity and photon index are similar to those characterizing the giant lobes of Centaurus A (labelled in Figure \ref{f2}).   
 
Stawarz at al. (2008) proposed a dynamical-radiative model of young radio sources, where the \g-ray emission is produced via inverse-Comptonization of circumnuclear (IR--to--UV) photon fields off relativistic electrons in  compact, expanding lobes. The level of the non-thermal high-energy emission predicted by the model depends on several source  parameters, such as the kinetic jet power, the accretion disk luminosity, and the lobes' compactness: sources with linear sizes $<$100 pc and  jet powers of $\gtrsim10^{46}$ erg s$^{-1}$ could reach \g-ray luminosities of the order of 10$^{44}$  erg s$^{-1}$ or larger (see also Orienti et al. 2011; D'Ammando et al. 2016). 
 The CSO \pkssev~represents an ideal candidate to test this model, due to its linear size of only $\sim 2$ pc and proximity (luminosity distance of 60.4 Mpc). 
 The symmetric radio morphology of \pkssev~suggests that we are observing the source at a large inclination angle, ruling out a \g-ray contribution of the boosted jet emission. The steady flux detected with {\it Fermi}-LAT is consistent with the \g-ray emission being isotropic and produced in the radio lobes expanding at a sub-relativistic velocity (Giroletti \& Polatidis 2009 report a hot spot expansion velocity of $\sim$0.07$c$). 

A detection in \g-rays may help us establishing the nature of the X-ray emission of CSOs (Tengstrand at al. 2009, Siemiginowska et al. 2016). At the angular resolution of the current X-ray observatories, this X-ray emission is usually spatially unresolved and could be a superposition of several distinct components, including that of the disk-corona system  and IC emission of the infrared photons of the putative torus off the lobes' electrons (Ostorero et al. 2010).
However, a proper test of the above models requires a comparison of the predicted broad-band spectral energy distribution with the available multiwavelength data (see Ostorero et al. 2010, Migliori et al. 2012). This is beyond the scope of this Letter and will be discussed in a forthcoming paper.

Is \pkssev~an isolated case or could we expect an increase of the detections of CSOs in \g-rays?  So far, \g-ray searches of young radio sources have not provided other significant detections (Migliori et al. 2014, D'Ammando et al. 2016). 
The 3FGL contains  three \g-ray sources tentatively identified as Compact Steep Spectrum (CSS) radio sources  (i.e. sources which likely are more evolved than CSOs): 3C\,84 (Kataoka et al. 2010; Dutson et al. 2014), 3C\,286, and 4C\,+39.23B (whose \g-ray association is however doubtful).
These are  recurrent or restarted radio sources with complex radio morphologies and multiple pairs of lobes on various linear scales.
A CSO classification has been proposed for three other \g-ray sources: 4C$+$55.17
(z=0.896, McConville et al. 2011), PMN\,J1603$-$4904 (z=0.18, M{\"u}ller et al. 2014, 2015), PKS\,1413+135 (z=0.247, Gugliucci et al. 2005). They show the evidence of a CSO-like morphology of the inner radio structures and the absence of the \g-ray and radio variability typically observed in blazars.
Differently from \pkssev, all these sources are located at relatively high redshifts, and display high  \g-rays fluxes with hard spectra. If their \g-ray emission is produced in the lobes and is thus isotropic, the mechanism producing it must be extremely efficient.

Other `\g-ray emitting CSOs' might be hiding among the large number of unidentified 3FGL \g-ray sources. If this were the case, it would be crucial to define an efficient strategy to unveil them.
The example of \pkssev~suggests that a detection in \g-rays is the most feasible for the most compact and nearby CSOs. On the other hand, no detection was reported for another very compact and nearby (z=0.076) CSO, OQ\,208  (Orienti et al. 2011, D'Ammando et al. 2016), implying that additional key-parameters may play a role. However, given its
redshift, OQ 208 would appear a factor of $\sim$30 fainter
than \pkssev, if its intrinsic \g-ray luminosity were the same as of \pkssev.

\acknowledgments
We are grateful to the anonymous referee for comments which helped to improve the paper. The authors thank P. Grandi for sharing the data of Figure \ref{f2}, and J. Ballet, F. D'Ammando and F. Massaro for useful suggestions.
G. M. and S. C. acknowledge the financial support from the UnivEarthS Labex program of
Sorbonne Paris Cit\'e (ANR­10­LABX­0023 and ANR­11­IDEX­0005­02). {\L}. S. was supported by Polish NSC grant DEC-2012/04/A/ST9/00083. 
L.O. acknowledges the grants: INFN InDark, MIUR 
PRIN2012 ``Fisica Astroparticellare Teorica'', and 
``Origin and Detection of Galactic and Extragalactic 
Cosmic Rays'' from UniTo and Compagnia di San Paolo.
This research is funded in part by NASA contract NAS8-39073. Partial support was provided by the Chandra grants GO4-15099X and GO0-11133X.

\clearpage

\begin{table}
\caption{ \pkssev~-- LAT Binned Analysis Results}
\label{t1}
\begin{center}
\medskip
\begin{tabular}{lccccc}
\hline
\hline

Energy Band   &Time             &TS            &$\Gamma$        & {\it Fermi}-LAT Flux                             &Log($\nu$F$_{\nu, {\rm 1GeV}}$)         \\
(1)                      &(2)            &(3)             &(4)                                       &(5)                                                                &(6)                        \\                           
\hline
\multicolumn{6}{c}{7-year dataset}\\
\\
0.1-100            &54686.49-57242.49           &36                 &2.9\er0.3                            &11.5\er0.3        &-12.4\er0.1                           \\

0.2-100            &54686.49-57242.49           &18.5             &2.6\er0.3                              &2.1\er0.8      &-12.4\er0.2                             \\
\hline
\multicolumn{6}{c}{Incremental Analysis}\\
\\
0.1-100            &54686.49-56147.49                    &15       &2.9(f)              &9.8\er2.8                 &-12.4\er0.13                                     \\
0.1-100            &54686.49-56512.49                    &19        &2.9(f)          &9.8\er2.5                   &-12.4\er0.11                                 \\
0.1-100            &54686.49-56877.49                    &28        &2.9(f)          &10.9\er2.3                  &-12.4\er0.09                                 \\
\hline
\end{tabular}
\end{center}
Columns: 1 -- Energy band selected for the analysis in GeV; 2 -- Observing time in Modified Julian Day; 3 -- Test statistic value; 4 -- Gamma-ray photon spectral index, (f) indicates fixed $\Gamma$; 5 -- {\it Fermi}--LAT  photon flux in the selected energy band in units of $\times10^{-9}$ ph cm$^{-2}$ s$^{-1}$; 6 -- Logarithm of the flux density at 1 GeV in erg cm$^{-2}$ s$^{-1}$.\\ 
\end{table}

\begin{table}
\caption{ \pkssev~ Gamma-ray best fit position}
\label{t2}
\begin{center}
\medskip
\begin{tabular}{lccc}
\hline
\hline
Source                                &RA                              &DEC                &Position uncertainty \\
      
(1)                                       &(2)                               &(3)                   &(4) \\ 
\hline
\\
\multicolumn{4}{c}{$\gamma$-ray position}\\
\\
3FGL\,J1728.0$-$6446      &17 28 02.29   &$-$64 46 23.08    &0.23$\times$0.20$^a$ (P.A.=79.5\dg)\\
                                           &                      &                            &0.37$\times$0.32$^b$ (P.A.=79.5\dg)\\
{\it gtfindsrc} best fit     &17 22 57.60  &$-$64 54 24.48    &0.18$^c$\\
position                              &                      &                            &0.31$^d$\\

\\
\multicolumn{4}{c}{radio position}\\
\\
\pkssev                               &17 23 41.0     &$-$65 00 36.6     &0.0020$\times$0.00095$^e$\\

\\
\hline
\end{tabular}
\end{center}
Columns: 1 -- Source; 2 \& 3 -- Source coordinates (RA in hh mm ss.d and DEC in dd mm ss.d); 4 -- Position uncertainty. \\ 
Notes: $^a$ \& $^b$ -- semi-major and semi-minor axes in degrees of the ellipse uncertainty region at 68\% and 95\% confidence level respectively, followed by the position angle of the 95\% confidence region (reported in the 3FGL); $^c$ \& $^d$ --  uncertainty radius of the best fit {\it gtfindsrc} position at 68\% and 95\% confidence level in degrees; $^e$ -- uncertainty on the radio position in arcseconds (Johnston et al. 1995).
\end{table}

\end{document}